\documentclass[epj]{svjour}
\usepackage{epsfig}

\begin{document}


\author{ M.Kiselev$^{1,2,*}$, F.Bouis$^1$, F.Onufrieva$^1$ and P.Pfeuty$^1$}
\institute{$^1$Laboratoire Leon Brillouin, CE-Saclay 91191,
Gif-sur-Yvette, France\\ $^2$ Russian Research Center "Kurchatov
Institute", 123 182 Moscow, Russia}
\date{\today}

\title{Some aspects of
electronic topological transition in 2D system on a square lattice.
Excitonic ordered states.}
\abstract{ We study the ordered "excitonic" states which develop
around the quantum critical point (QCP) associated with the
electronic topological transition (ETT) in a 2D electron system on
a square lattice. We consider the case of hopping beyond nearest
neighbors when ETT  has an unusual character. We show that the amplitude 
of the order parameter (OP) and
of the gap in the electron spectrum increase with increasing the
distance from the QCP, $\delta_c - \delta$, where $\delta = 1-n$  
and $n$ is an electron concentration.
Such a behavior is different from the ordinary
case when OP and the gap decrease when going away from the point which is a
motor for instability. The gap opens at "hot spots" and extends  
untill the saddle points (SP) whatever is the doping concentration. The
spectrum gets a characteristic flat shape as a result of hybrydization effect
in the vicinity of two different SP's. 
The shape of the
spectrum and the angle dependence of the gap have a striking
similarity with the features observed in the normal state  of the
underdoped high-T$_c$ cuprates. We discuss also  details about the
phase diagram and the behaviour of the density of states.}
\titlerunning{Some aspects of ...}
\maketitle

 Many experiments performed for high $T_c$
cuprates provide an evidence for the existence of a pseudogap in
the underdoped regime above $T_c$ and below some temperature
$T^*(\delta)$ which value increases with increasing the doping
distance from the optimal doping, $\delta_{opt}-\delta$
\cite{NMR}-\cite{Raman}. The pseudogap is observed directly by
angle-resolved photoemission spectroscopy (ARPES) measurements
\cite{ARPES}-\cite{ARPES4}. The striking
 about this gap is its increase with increasing
$\delta_{opt}-\delta$ \cite{ARPES2} while the critical temperature
of superconducting (SC) transition, $T_{sc}$, decreases. Another
prominent  feature is the so-called $(\pi,0)$ feature discovered
by ARPES: the electron spectrum around  the saddle-point (SP)  is
flat and disappears above some threshold value of
wavevector  \cite{ARPES}. Several hypothesis exist about possible
origin of the pseudogap \cite{KaS}-\cite{Alt}. In this paper we
present another explanation of this phenomenon in the framework of
the model developed in  \cite{PRL2}-\cite{PRL1}. In these works
the concept of the Electronic Topological Transition in 2D system
is developed and applied for the explanation of various effects
experimentally observed in High - T$_c$ cuprates.

In the present paper we consider various ordered states appearing
in the vicinity of ETT point. We show that the ordered "excitonic"
phase \cite{On2} formed in a proximity of QCP
 in a 2D fermion system
on a square lattice  is characterized (on
one side of it, $\delta<\delta_c$) by the electron spectrum
strikingly similar to that observed in the underdoped cuprates.
The mentioned QCP corresponds to the electron concentration
$n_c=1-\delta_c$ at which Fermi level (FL) crosses saddle points
(SP) in the bare spectrum. As shown in \cite{On2}, in the case of
hopping between more than nearest neighbors this point is a point
of a fundamental electronic topological transition (ETT)  for which
singularities in thermodynamical properties (and the logarithmic
 divergence in density of states at $\omega=0$) is only first quite trivial
aspect (related to the local change in topology of the Fermi
surface (FS)). The other aspects of criticality related to the mutual change in
topology of FS near two different SP's lead to a very asymmetric
behaviour of the noninteracting and interacting system on two
sides of ETT being quite anomalous on the side $\delta<\delta_c$.
On the other hand, for realistic for the high-$T_c$ cuprates
ratios of hopping parameters $t'/t$, $\delta_c$ is given by :
 $\delta_c=0.27$ for $t'/t=-0.3$ and $\delta_c=0.17$ for $t'/t=-0.2$,
i.e. the anomalous regime $\delta<\delta_c$ occurs in the doping
 range   where the experimentally observed
"strange metal" behaviour takes place. Moreover,
 $\delta=\delta_c$ corresponds to a maximum of
$T_{sc}(\delta)$ (as discussed in \cite{On2}) and therefore the latter regime
 can be considered as an underdoped regime.

Some  anomalies concerning the ordered "excitonic" phase have been
discussed in \cite{PRL2}-\cite{PRL3}. Namely, it was
shown that the line of the "excitonic" instability
  grows from QCP
to the side
$\delta<\delta_c$
 instead of having
the form of a bell around QCP as it usually happens for an
ordinary QCP. Other anomalies which exists in the ordered phase
are considered in the present paper. [We call this  phase
"excitonic" ordered phase because the discussed instability has
the same origin as the classical "excitonic" instability
intensively discussed in the 60-70 \cite{KeK}-\cite{Rice}. Namely
it is related to the opposite curvature of two parts of  electron
spectrum in a proximity of FL. In the case considered they
correspond to spectra in vicinities of two SP's.]

We consider various possibilities for the ordered state, namely,
Spin and Charge Density Wave orderings with different types of the
order parameter symmetries (s-wave, d-wave) depending on the
effective interaction between the quasiparticles. Despite of the
different symmetries, the properties of such ordered states
resemble an "excitonic" states \cite{KeK}-\cite{Rice} are quite
similar. The explanation of this phenomenon is in the fundamental
role of ETT and  effects of criticality in the vicinity of the
corresponding QCP \cite{PRL1}. We show that the electron spectrum
in the ordered phase is characterized by a gap on FL which opens
at "hot spot" and extends until SP whatever is the doping
concentration. Therefore there is always a gap at the SP
wavevectors $(0,\pm\pi)$, $(\pm\pi,0)$. This remarkable feature is
related, as we show in the paper, to a quite nontrivial aspect of
ETT: it is the end point of two critical lines for the
 "polarization operator" characterizing a behaviour of
the free electron system. The other side of the same effect
 is an increase of the amplitude of the
 order parameter (and of the  gap)
with increasing the distance from QCP on the underdoped side. We
show also that the electron spectrum in a vicinity of SP  gets a
specific "flat"   form which on one hand is typical for an
"excitonic" phase (see for example \cite{Halp1}) being a result of
a  hybridization of two parts of the bare spectrum with the
opposite curvature and on the other hand  has a striking
similarity with the form of the spectrum observed by ARPES
\cite{ARPES}-\cite{ARPES4}.
 We show  that the spectrum
"disappears" above some threshold value of wavevector in the
direction $(\pi,0)-(\pi,\pi)$ that is also an effect  of the same
hybridization. We briefly discuss also features related to
strong-coupling limit of the model and effects of strong electron
correlations.

A starting point is a 2D system of free fermions
on a square lattice
 with hopping between   nearest ($t$) and  next
nearest ($t'$) neighbors

\begin{equation}
\epsilon_{{\bf k}} = \ -2t (\cos k_x + \cos k_y ) - 4t' \cos k_x
\cos k_y \label{1}
\end{equation}
 (as in \cite{On2} we consider $t>0$ and $t'<0$) and $|t'/t|$ not too
small and not too close to the limit $|t'/t|=.5$.
  The dispersion law (\ref{1}) is characterized
 by two different saddle points (SP's) located at
 $ ( \pm\ \pi, 0)$ and $(0, \pm  \pi)$
  (in the first Brillouin zone $ (-\pi,0)$ is equivalent to $ (\pi,0)$ and
$ (0,-\pi)$ is equivalent to $ (0,\pi )$)  with the energy
$\epsilon_{s} = 4 t'$ .
When we vary the chemical potential $\mu$ or the energy
 distance from
the SP, $Z$, determined as

\begin{equation}
Z=\mu-\epsilon_{s}=\epsilon_{F}-4 t',
\label{3}
\end{equation}
 the topology
of the Fermi surface changes when $Z$ goes from $Z<0$ to $Z>0$ through
the critical value $Z=0$, see Fig.\ref{f1}.

In \cite{On2} we have
 shown  that such a system
undergoes a fundamental ETT at the electron
concentration corresponding to $Z=0$.
The corresponding quantum critical
point is  quite rich. It combines several  aspects
of criticality. The first standard one is related to
 singularities in thermodynamic
properties, in  density of states at $\omega=0$
(Van Hove singularity),
to  additional singularity in the
  superconducting
(SC) response function, all reflect a local
change in the topology of FS.
This aspect is not important
for the properties we are interested in the present
paper. Important aspects which reflect a mutual
change in the topology of FS in the vicinities of
two SP's are the following.
First of all, it is a logarithmic divergence
of the
polarizalibility of
noninteracting electrons

\begin{equation}
\chi^0({\bf k},\omega)=\frac{1}{N}\sum_{{\bf q}}\frac{n^{F}(\tilde{\epsilon}_{{\bf q}})
-n^{F}(\tilde{\epsilon}_{{\bf {q+k}}})}{\tilde{\epsilon}_{{\bf q+k}
}-
\tilde{\epsilon}_{{\bf {q}}}-\omega-i0^+},
\label{11}
\end{equation}
 as ${\bf k}={\bf Q}=(\pi,\pi)$,
 $\omega = 0$ and   $Z  \rightarrow  0$ :

\begin{equation}
\chi^{0}({\bf Q},0) \propto \ln\frac{\omega_{max}}{|Z|}.
\label{13}
\end{equation}
which has an "excitonic" origin ($\omega_{max} \sim t$ is a cutoff
energy). By "excitonic" origin we mean that two branches of the
spectrum corresponding to vicinities of two SP's $(a=t-2t', \hskip
0.3 cm b=t+2t')$

$$ \tilde{\epsilon}_1({\bf k}) = \epsilon_1({\bf k})-\mu= -Z+a
k_x^2 - b k_y^2,$$
\begin{equation}
\tilde{\epsilon}_2({\bf k}) =\epsilon_2({\bf k})-\mu = -Z+a k_y^2
- b k_x^2 \label{16}
\end{equation}
have such a form (see Fig.\ref{f2}) that
 at $Z=0$ the chemical
potential lies on the bottom of one "band"
and on the top of the another  for the
given directions
$(0,\pi)-(\pi,\pi)$ and $(\pi,0)-(0,0)$, (see Fig.\ref{f2}).
 Therefore, no energy is needed
to excite the electron-hole pair. It is this divergence
 that is at the origin of density wave (DW) instability. 
 The DW instability can be of 
 Spin Density Wave (SDW), Charge
Density Wave (CDW), Spin Current Density Wave (SCDW) or Orbital
Current Density Wave (OCDW) instability \cite{Tug}) of interacting
electron system depending on a nature of interaction.

 The nontriviality
stems from the  aspect of criticality related
to the effect of Kohn singularity in 2D system :
the point  $Z=0$, $T=0$ is  the end
of the critical line  $Z<0$ each point of which is
a point
of   static
Kohn singularities in polarizability of noninteracting
electrons. As shown in \cite{On2},
the latter aspect is a motor for the anomalous behaviour
of the system on the  side $Z>0$ of ETT. One among the
anomalies
found in \cite{On2} concerns
 the ordered DW phases. We have obtained that
the line of DW  "excitonic"
instability $T_{DW}(Z)$
 has
the anomalous
 form  on the side $Z>0$ : it grows from QCP
instead of having the form of a bell
around QCP as it usually happens in the case of ordinary QCP
and as it indeed happens on the side $Z<0$.
 Below we show that this latter aspect
 is also at the origin of anomalous behaviour
of the order parameter and of some other anomalies
 in the ordered state in the same regime $Z>0$.

As shown in \cite{On2}, on the side $Z>0$  of the electronic
topological transition, a maximum of the static electron-hole
susceptibility occurs at the wavevector ${\bf q=Q}$.
Therefore in a presence of $\bf q$ independent interaction
or $\bf q$ dependent interaction negative for ${\bf q=Q}$,
the DW  instability happens at
 ${\bf q=Q}$ and this is
the wavevector of ordering
 in the  DW phase.
As usual for such phases, one should consider a matrix
 electron Green function
 containing as components the normal and
anomalous Green functions in terms of operators
$a^+_{{\bf k},\sigma}$ and $a_{{\bf k}\sigma}$ which are the creation and
annihilation electron's operators respectively:

$$
K_{11}({\bf k},i\omega_n)=-\int_0^\beta d\tau e^{i\omega_n \tau}
<T_\tau a_{{\bf k}\sigma}
(\tau)|a^+_{{\bf k}\sigma}(0)>$$
$$ K_{22}({\bf
k},i\omega_n)=-\int_0^\beta d\tau e^{i\omega_n \tau}
<T_\tau a_{{\bf k+Q}\sigma}
(\tau)|a^+_{{\bf k+Q}\sigma}(0)>
$$

\begin{equation}
K_{12}^{\sigma\sigma'}({\bf k},i\omega_n)=-\int_0^\beta d\tau e^{i\omega_n \tau}
<T_\tau
 a_{{\bf k+Q}\sigma}
(\tau)|a^+_{{\bf k}\sigma'}(0)>.
\label{1.10}
\end{equation}
[Below we will omit spin indices in the Green functions keeping in
mind that $K_{12}=K_{12}^{\sigma -\sigma}$ for CDW and OCDW states
and $K_{12}=K_{12}^{\sigma\sigma}$ for SDW and SCDW states.]

If the anomalous Green function   $K_{12}$
is nonzero (that should be found selfconsistently)
the explicit expressions for the two Green functions are
as follows

$$ K_{11}({\bf k},i\omega_n)= \left[\frac{u^2({\bf k})} {i\omega_n
- \varepsilon_1} +\frac{v^2({\bf k})} {i\omega_n - \varepsilon_2}
\right]$$
 $$K_{22}({\bf k},i\omega_n)= \left[\frac{u^2({\bf k})}
{i\omega_n - \varepsilon_2} +\frac{v^2({\bf k})} {i\omega_n -
\varepsilon_1}, \right] $$
 $$K_{12}({\bf k},i\omega_n)=
K_{21}({\bf k},i\omega_n)=$$
\begin{equation}
= u({\bf k})v({\bf k})\left[\frac{1}
{i\omega_n - \varepsilon_1} -\frac{1}{i\omega_n - \varepsilon_2}\right],
\label{1.20}
\end{equation}
where $u$, $v$ - coefficients have a standard form :

$$
 u^2({\bf k})  = \frac{1}{2}\left[1+
\frac{\epsilon_A(k)-\epsilon_B(k)}{2E({\bf k})}\right],$$
$$
v^2({\bf k})  = \frac{1}{2}\left[1-
\frac{\epsilon_A(k)-\epsilon_B(k)}{2E({\bf k})}\right], $$
\begin{equation}
E({\bf k})  =  \sqrt{\left(\frac{\epsilon_A-\epsilon_B}{2}\right)^2+
|\Delta({\bf k})|^2}.
\label{1.21}
\end{equation}
The  spectrum in the ordered state is given by

$$
\varepsilon_{1,2}=\frac{\epsilon_A+\epsilon_B}{2}
\pm
\sqrt{\left(\frac{\epsilon_A-\epsilon_B}{2}\right)^2+
|\Delta({\bf k})|^2},\;\;\;\;\;\;\;
$$
\begin{equation}
\epsilon_A({\bf k}) \equiv \epsilon({\bf k})\;\;\;\;\;\;\;\;
\epsilon_B({\bf k})=\epsilon({\bf k+Q}),
\label{1.19}
\end{equation}
where $\epsilon({\bf k})$ is defined by (\ref{1}).
The equation for the gap
is

\begin{equation}
\Delta({\bf k})= -T\sum_{\omega_n}\frac{1}{N}\sum_{\bf p}
\Gamma_{12}({\bf k, k+Q, p})K_{12}
({\bf p},i\omega_n)
\label{1.18}
\end{equation}
where $\Gamma_{12}$ is a vertex which in mean field approximation
coincides with the bare interaction:
\begin{eqnarray}
\nonumber \Gamma_{12}({\bf k,k+Q,p})= V_{\bf Q},\;\;\; for\;
SDW\;\;(CDW)\\ \Gamma_{12}({\bf k,k+Q,p})= V_{\bf k-p},\;
for\;OCDW\;\;(SCDW) \label{Gam}
\end{eqnarray}
where $V_{\bf k}=2V(\cos(k_x)+\cos(k_y))$.
 The type of
the interaction and therefore, type of the excitonic phase depend on
the model.  The SDW and OCDW instabilities occur
in the case of a positive interaction in the triplet
channel (exchange interaction), the CDW and SCDW instabilities 
take place for positive interaction in the singlet
channel (density-density interaction).
 We will not fix for the moment a type
of interaction and therefore a nature of the ordered phase
assuming that there exists either the first or the second
interaction.

The equation (\ref{1.18}) is reduced to the following equation

\begin{equation}
1=4|V|\Pi^{DW}_{\bf k=0}({\bf Q},Z,\Delta)
\label{Pi}
\end{equation}
where the "polarization operators"   $\Pi^{DW}({\bf
Q},Z,\Delta)$ are given by one of the following equations \cite{DW}:
$$\Pi^{SDW, CDW}({\bf Q},Z,\Delta)=$$
\begin{equation}
=\frac{1}{4N}\sum_p\frac{1}{E({\bf p})}
[\tanh(\frac{\varepsilon_1}{2T})-\tanh(\frac{\varepsilon_2}{2T})],
\label{pidw}
\end{equation}

$$\Pi^{OCDW,SCDW}({\bf Q},Z,\Delta)=$$
\begin{equation}
=\frac{1}{4N}\sum_p \frac{(\cos p_x-\cos p_y)^2}{4E({\bf p})}
[\tanh(\frac{\varepsilon_1}{2T})-\tanh(\frac{\varepsilon_2}{2T})]
\label{GE}
\end{equation}
The  expressions (\ref{Pi}) are the equation for the SDW, CDW, OCDW or SCDW
gap which should be solved selfconsistently. We emphasis that for $V>0$ only
SDW (OCDW) solution is possible whereas  CDW (SCDW) solution takes place
for $V<0$.

The solution of (\ref{Pi})-(\ref{GE}) is given by 
one of the following expressions
$$\Delta=\Delta^{SDW}({\bf k})=\Delta^{SDW}_0,$$
$$\Delta=\Delta^{CDW}({\bf k})=\Delta^{CDW}_0,$$
$$\Delta=\Delta^{OCDW}({\bf k})=\Delta^{OCDW}_0 (\cos k_x -\cos k_y)/2,$$
\begin{equation}
\Delta=\Delta^{SCDW}({\bf k})=\Delta^{SCDW}_0 (\cos k_x -\cos k_y)/2
\label{Dlt}
\end{equation}

The equations (\ref{1.18})-(\ref{Dlt}) are quite standard. A nontriviality,
as we show below,
is related to the behaviour of the "polarization
operator"
 in a
proximity of ETT.
As we have shown in \cite{On2}, the effect that the point of ETT
 is the end point of the critical line $Z<0$
leads to the anomalous behaviour
of the  electron-hole susceptibility
$\chi^0({\bf Q},Z,\omega)$
on the side $Z>0$. Below we show that a similar
effect takes place for the
"polarization
operator" (\ref{GE}). The two functions coincides in the
 limit
cases:
$\chi^0({\bf Q},Z,\omega=0)=\Pi^{DW}({\bf Q},Z,\Delta=0)$.
[It is important to emphasize that
the behaviour of the "polarization operator"
depends  only on properties of
the system of noninteracting electrons, namely on the
topology of FS.]

Calculated for $T=0$ "polarization
operators"
 $\Pi^{DW}$ \newline
 $({\bf Q},Z,\Delta_0)$
as a function of $\Delta_0$ for fixed $Z$ (in the regime $Z>0$)
 are shown in Fig.\ref{f3}. Since the properties of the "polarization operators"
are similar in many aspects we shall omit later the indices (SDW, CDW, OCDW or SCDW) except for the cases when it will be necessary to emphasize the difference.

One can see that there is a singularity  at some point
 $\Delta_0=\Delta_c(Z)$. The value of $\Delta_c(Z)$
increases with increasing $Z$. The situation is quite
similar to that analyzed in \cite{On2} for $\chi^0$ as a function of $\omega$
for fixed $Z$  and
$T=0$. In the latter case we have found a square-root singularity
at
$$
\omega= \omega_c=\frac{2Z}{1-2t'/t},
$$
which is the dynamic Kohn singularity. As we see in Fig.3,
for the polarization operator $\Pi({\bf Q},Z,\Delta)$
 the singularity is weaker,
while $\Delta_c(Z)$ also scales with $Z$.

Analytical estimations show  that
$\Delta_c(Z)$ is given by

\begin{equation}
\Delta_c(Z)=Z
\label{DC}
\end{equation}
while the asymptotic form of $\Pi(Q,Z,\Delta)$ near the
singularity is given by  :

\begin{equation}
t \Pi(Q,Z,\Delta)=
 \left\{
\begin{array}{cc}\displaystyle
A_1 |1-\Delta/\Delta_c| +B,
& \hskip 0.3 cm \Delta<\Delta_c\\
\displaystyle
A_2 |1-\Delta/\Delta_c| +B,
& \hskip 0.3 cm \Delta>\Delta_c.
\end{array}
\right.
\label{tPi}
\end{equation}
The jump in the derivative, $A_1-A_2$, depends only
on $t'/t$ \cite{const} and is proportional to 

\begin{equation}
A_1-A_2 \propto \frac{1}{|t'/t|}
\left(\ln|\frac{4}{t'/t}| - A_0\right),
\label{A12}
\end{equation}
where $A_0$ is a constant (for the spectrum (\ref{16}) $A_0=\pi/8$).
 The critical line (\ref{DC}) is clearly seen in Fig.4a where we
present the calculated $\Pi({\bf Q},Z,\Delta_0)$ as a function of
$Z$ and $\Delta_0$. From the point of view of the behaviour of the
"polarization operator", the ETT point is the end point of two
critical lines. The first is the semiaxis $Z<0$ each point of
which corresponds to the square-root singularity in $\Pi({\bf
q},Z,\Delta)$ occurring as $\Delta_0 \rightarrow 0$ and ${\bf q}
\rightarrow {\bf q}_m$, where the latter is the characteristic for
this regime wavevector of incommensurability (see \cite{On2} where
the ${\bf q}$ dependence of $\Pi({\bf q},Z,0)$= $\chi^0({\bf
q},Z)$ is analyzed in details.) The second is the line
$Z=\Delta_0$ each point of which corresponds to the kink in
$\Pi({\bf q},Z,\Delta(q))$ occurring at $T=0$ as ${\bf q}
\rightarrow {\bf Q}$  where the latter is the characteristic
wavevector for the regime $Z>0$. At the point of intersection of
these lines, $Z=0$, the two types of singularities are
transforming into the logarithmic singularity ;  $\Pi({\bf
q},0,\Delta) \propto \ln|\max({\bf q-Q},\Delta)|$.

The existence of the growing with $Z$ critical line determines
a quite unusual form of the lines
 $\Pi(Q,Z,\Delta)=constant$ which
develop around the critical line $\Delta_c(Z)$ and grow
with increasing $Z$,  (see Fig.\ref{f4}(b)).

In preceeding discussion we presented some general analysis which does not depend on
details of interaction considered but only on the topology of the FS.
To provide the calculations, let us consider a particular case of interaction
resulting in Spin Density Wave (\ref{Gam}),(\ref{pidw}). 
The solution of corresponding eq. (\ref{Pi}) for $t/V=1.8$
is shown in Fig.\ref{f5a}. Two branches of the
solution have an anomalous dependence of the gap on $Z$
reproducing the form of the lines  $\Pi(Q,Z,\Delta_0)=constant$ in
Fig.\ref{f4}(b). The anomaly is that for both solutions {\bf gap
increases with increasing the distance from the quantum critical
point}, i.e. from the point which is at the origin of the ordered
phase. [For an ordinary QCP the gap is maximum at the electron
concentration corresponding to QCP and decreases monotonously with
increasing the distance from QCP. For example such a picture takes
place for DW phase on both sides from QCP in the case of
$t'=t''=...=0$; as we discussed in \cite{On2} in the latter case
all anomalies in the regime $\delta<\delta_c$ disappear. In the
case considered in the paper it happens on the overdoped side of
the QCP.]

The difference between two solutions for the gap
presented in Fig.\ref{f5a} is that

\begin{equation}
\Delta_1(Z)>Z
\label{DgZ}
\end{equation}
while
\begin{equation}
\Delta_2(Z)<Z
\label{DmZ}
\end{equation}
for any $Z$, any $t/V$, any $t'/t$ since the two lines,
$\Delta_1(Z)$ and $\Delta_2(Z)$
are attached to the critical line $\Delta=\Delta_c(Z)=Z$ from above
and from below. For the most range of the existence of the
ordered phase $Z<Z_{cr}^{(1)}$,
 see Fig.\ref{f5a},
only one solution exists, the one corresponding to
eq. (\ref{Pi}). In the hyperbolic
approximation and under the condition $|t'/t|$ not too small $Z_{cr}^{(1)}$
is given by:
$$Z_{cr}^{(1)} \propto \omega_{max}\exp(-\pi^2t/(V\ln|t/t'|)).$$
\noindent
For this solution one has

\begin{equation}
\Delta_1(Z)\equiv\Delta_0(Z) = f(Z)+ \Delta(0)
\label{DfZ}
\end{equation}
where $\Delta(0)$ is given by

\begin{equation}
\Delta(0) \propto |t'|\exp(-\frac{2\pi^2 |t'/V|}{\sqrt{1-(2t'/t)^2}}),
\label{D0}
\end{equation}
and $f(Z)$ is an increasing function of $Z$, linear under
the condition
 $\Delta(Z) \gg \Delta(0)$.
The expression (\ref{D0}) is valid under condition
$\pi^2 |t'/V|/\sqrt{1-(2t'/t)^2} \gg 1$.
For the narrow $Z$ range of the coexistence
of the two solutions $Z_{cr}^{(1)}<Z<Z_{cr}^{(2)}$ it is the solution $\Delta_1$
 which is favorable (see Appendix).
{\bf Therefore, the value of the gap
increases with increasing $Z$ being always
 larger than $Z$.} As we have shown, this is a consequence
of the effect that the point of ETT is the end point of
 two critical lines.

Let's analyze now the form of the
spectrum in the DW phase. The spectrum given by (\ref{1.19})
is plotted in Fig.\ref{f6}.
for three important directions : $(\pi,\pi)-(\pi,0)-(0,0)$
and $(0,0)-(\pi,\pi)$.
The spectrum in
the vicinity of SP has the following prominent features:
The first is a characteristic "flat" shape (very close to the
experimental shape \cite{ARPES}, see Fig.\ref{f7}(a)
being a consequence of the
hybridization of the two branches of the bare spectrum
in the vicinity of two different SP's
with the opposite
curvatures, (see Fig.\ref{f5}).

The second: the spectrum in the direction $(\pi,\pi)-(0,\pi)$
"disappears" above some threshold value of wavevector since the
residue $v^2_{\bf k}$ tends to zero (that is also an ordinary
consequence of the hybridization). On the other hand, since
$$\varepsilon_1({\bf k}_{SP})=-Z+\Delta,$$
$$\varepsilon_2({\bf k}_{SP})=-Z-\Delta $$
(see, e.g. Eq.(\ref{1.19})) and $\Delta > Z$,
{\bf the chemical potential always lies in
the gap} for the part of Brillouin zone (BZ) starting from the
"hot spot" {\bf until SP} that is a consequence of the existence
of the critical line $\Delta=\Delta_c$ related to the discussed above
aspect of criticality of the QCP. For the direction $(0,0)-(1,1)$,
Fermi level lies in the lower branch of the spectrum, (see
Fig.\ref{f6}(b)), i.e. the system remains metallic. The
theoretical spectrum has a striking similarity with the anomalous
experimental
 electron spectrum in
the underdoped cuprates observed below the
characteristic line $T^*(\delta)$
  by ARPES \cite{ARPES},  we reproduce it in Fig.\ref{f7}. We remind that
ARPES measures a spectral function only below FL.

Then in Fig.\ref{f8} we present  the angle dependence of the value of
$\epsilon_{\bf k}-\mu$, i.e. of the
gap {\bf calculated from FL}, in the same way
as it is done in ARPES experiments \cite{ARPES0}.
Namely we plot the minimal value of $|\varepsilon_{\bf k} - \mu|$
for each given direction.
 The dependence is of a "d-wave type" in a sense that
the gap increases with increasing the argument
$(\cos k_x-\cos k_y)$ almost linearly in the proximity
of SP. However the dependence
 is flat (not linear as it happens in the d-wave case)  when approaching the direction $(1,1)$. Such a behaviour
 is also close to the experimentally found behaviour
above $T_c$ \cite{ARPES0} reproduced in Fig.\ref{f8}(b).
[Although the authors of \cite{ARPES0}
claim that the behaviour observed above and below $T_c$ is the same,
 what one sees in the experimental plot is not exactly this :
the  behaviour above and below $T_c$ is similar in the vicinity
of SP and different when approaching the $(1,1)$ direction
and this occurs quite systematically, see also the plots in \cite{ARPES0}
for other samples.]

We considered the particular case of SDW as an example of ordreed
"excitonic" state. Nevertheless, all aforesaid is true for any other
types of ordered states since the existence of such states
is determined only by topology of FS.

Let us study now the one particle density of states (DOS)
 given by the expression
$$ {\cal N}_{}(\varepsilon) =-\frac{1}{\pi}\frac{1}{N} \sum_{\bf
p}[{\it Im} K_{11}^R({\bf p},\varepsilon) +{\it Im} K_{22}^R({\bf
p},\varepsilon)]=$$
\begin{equation}
=\frac{1}{N}\sum_{\bf p}[\delta(\varepsilon - \varepsilon_1({\bf
p}))+ \delta(\varepsilon - \varepsilon_2({\bf p}))], \label{1.331}
\end{equation}

Numerical calculations with the spectrum (\ref{1}) give the
picture shown in Fig.\ref{f9}.

 The density of states of SDW (CDW) states deviates
from the DOS in the initial metallic state in two
$\varepsilon$ ranges notated as {\bf A} and {\bf B}.
For OCDW (SCDW) states only feature  {\bf A} survives.
Analytical calculations show that the {\bf A}-feature
is related to the existence of the discussed above
QCP (which we call below QCP1).
 Calculations
of the integral in (\ref{1.331}) performing with the hyperbolic
spectrum (\ref{16}) valid
in the vicinities of SP's show
 that in the {\bf A} range
DOS is characterized by
three singularities (instead of one logarithmic singularity
in the bare density of states ${\cal N}_0(\varepsilon) $ as $\varepsilon \to -Z$).
Those are a logarithmic singularity at
$\varepsilon_1 = -Z-\Delta_0\sqrt{1-4(t'/t)^2}$
\cite{Arist}

\begin{equation}
{\cal N}(\varepsilon \to \varepsilon_1)  \sim \frac{1}{t\sqrt{1-4(t'/t)^2}}
\ln(\frac{\omega_{max}}{|\varepsilon - \varepsilon_1|})
\label{DOSA}
\end{equation}
and
jumps at two energies
$$
\varepsilon_{2,3} = -Z \pm \Delta_0.
$$
The distance between two jumps  is  equal to $2\Delta$.

The {\bf B} feature is related to the existence
of the second
quantum critical point in the system
(QCP2) discussed in \cite{On3}.
This point corresponds to the electron concentration
when the chemical potential is equal :
$\mu=\mu_{c2}=0$
or by other words  when
the wavevector connecting two parts of FS in the direction $(1,1)$
is equal to ${\bf Q}_{AF}=(\pi,\pi)$.
 In this case two "hot spots" on FS come
together at the singular position $(\pm\pi/2,\pm\pi/2)$
 before disappearing.
The calculations of the integral in (\ref{1.331}) with the spectrum taken around
$(\pi/2,\pi/2)$
give a logarithmic divergence at the point
$\varepsilon_4=-Z-4t'+\Delta(\pi/2,\pi/2)$:

\begin{equation}
{\cal N}(\varepsilon \to \varepsilon_4)
-{\cal N}_0(\varepsilon_4) \sim \frac{1}{t}\sqrt{\frac{\Delta(\pi/2,\pi/2)}{|t'|}}
\ln(\frac{\omega_{max}}{|\varepsilon - \varepsilon_4|})
\label{DOSB}
\end{equation}
and a  jump at the point $\varepsilon=-Z-4t'-\Delta(\pi/2,\pi/2)$. This feature
does not exist for OCDW (SCDW) states since $\Delta({\bf k})=0$ along
the diagonal of BZ.

The  {\bf B} feature is important in the case
 when the chemical potential lies close to
the pseudogap in the {\bf B} part that should take place
in the
electron-doped cuprates. For the hole-doped cuprates
we are interested in the present paper, it is QCP1 which determines
properties of the system. In this case
the chemical
potential lies in the "pseudo-gap" {\bf A} according to the  properties
of the electron spectrum in the vicinity of SP discussed above.

Let's analyze now the range of the existence of the ordered phase
in the $T-Z$ plane. For this sake let's analyze the behaviour of
$\Pi({\bf Q},Z,\Delta_0)$ as a function of $Z$ at finite
temperature. [We again consider SDW state for certainity.]
 Results of calculations are presented in Fig.\ref{f10}. The first
observation is that the gap changes only little with $T$ at low $T$. The
second is that the behaviour at finite temperature as a function
of $Z$  is qualitatively the same as for
$T=0$ and it is anomalous: {\bf the  value
of the gap increases with increasing $Z$}.

The phase diagram in $T-Z$ plane obtained for SDW  (CDW) instability
based on the
analysis of the gap behaviour at finite $T$ is presented
in Fig.\ref{f11}. It is worthwhile to note 
that the polarization operator
$\Pi$ $({\bf Q},Z,\Delta)$ (\ref{GE})
calculated for OCDW (SCDW) ordered states 
has essentially more abrupt behavior as
a function of $Z$  in comparison with
those for $\Pi^{SDW,CDW}$$({\bf Q},Z\Delta)$ (\ref{pidw}).
Such behaiour appears
due to  additional factor $(\cos(p_x)-\cos(p_y))^2$
in the integral (\ref{Pi}). As a result, the domain of existence 
of OCDW (SCDW) solutions for equation (\ref{Pi}) at various doping
concentrations is  substantially narrower than for SDW (CDW) case.
Nevertheless, it does not affect on the qualitative shape of 
phase diagram Fig.\ref{f11}.

The solid line is the line where $\Delta_1(T)=0$.
The dashed line is the line where $\Delta_2(T)=0$. These
two lines are at the same time the lines of instabilities of the
undistorted metallic state. The line $\Delta_2(T)=0$ is
not however a line of a phase transition since
the nonzero
 solutions for the gap exist on the left
of this line until the dot-dashed line. Along the latter line
corresponding to the disappearance of the "ordered" solution,
{\bf the gap is
finite} and the two solutions coincide : $\Delta_0(T)=
\Delta_1(T)=
\Delta_2(T)$.
The situation is clearly seen from Fig.\ref{f13} where we present the lines
$\Pi({\bf Q},Z,\Delta)=const$ for different $Z$ and
fixed $t/J$ which in fact give the full picture of the
behaviour of the DW gap as a function of $Z$ and $T$.

As we discuss in the Appendix, in the region between the dot-dashed and dashed line, where three solutions $\Delta_0=\Delta_1$, $\Delta_0=\Delta_2$ and $\Delta=0$
coexist, it is the solution $\Delta_0=\Delta_1$ which is energetically favorable.

Thus, the dot-dashed line in the phase diagram in
Fig.\ref{f11} is the line of the first-order
phase transition. The gap along this line changes only little
at low temperature and tends to zero rapidly in the vicinity of
the point $O$. The latter  is a tricritical point. The
range in $T-Z$ plane in the vicinity of this
 point  corresponds to a strongly fluctuating regime
which we will consider elsewhere. It is important to add also that
at the point $Z=Z_{cr}^{(2)}$ of the appearance of the ordered
phase at $T=0$, the gap is exactly equal to $Z$ that means that
the upper branch of the spectrum in Fig.\ref{f6}(a) touches FL.
Then when moving inside the ordered  phase the gap $\Delta$
becomes larger than $Z$ and this branch goes up leaving the FL.

Above we have considered the critical temperatures
and the gap behaviour as functions
of the energy distance from the QCP, $Z$. It is worth for applications
to cuprates to change the description and to consider
physical properties as functions of
 electron concentration $n_e$ or of hole doping $\delta=
1-n_e$.
To do this we  use the relation between $Z$
(or the chemical potential $\mu$) and the hole doping
which for $T=0$ is given by  :

\begin{equation}
1-\delta= \int_{\omega}N(\omega)d\omega.
\label{30}
\end{equation}
So far as
\begin{equation}
Z \propto \delta_{c}- \delta,
\label{31}
\end{equation}
all dependencies considered above can be rewritten as  functions
of doping distance from QCP. For example, the phase diagram in the
plane $T-\delta$ calculated for $t'/t=-0.3$ for which
$\delta_c=0.27$ gets the form shown in Fig.\ref{f12}.

One can easily obtain values of doping for all plots
presented in Fig.\ref{f4}-Fig.\ref{f10} when comparing the phase
diagram in  $T-Z$ plane in Fig.\ref{f11}, and
in  $T-\delta$ plane in Fig.\ref{f12}.

Obviously, the gap $\Delta_0(\delta)$ increases with
$\delta_c-\delta$ in the same way as it increases
with $Z$, see Fig.\ref{f5a} and Fig.\ref{f10} for $\Delta_0=\Delta_1$.


All features discussed above do not depend on the nature of the
ordered phase, SDW, CDW, OCDW or SCDW since they reflect the
topological aspects of ETT. The type of the excitopic phases
developing around ETT point depend on the
type of interaction. It is the SDW or OCDW state in the case of a
positive interaction in the triplet channel (exchange interaction)
and the CDW or SCDW state in the case of a positive interaction in
a singlet channel (density-density interaction).  The ordered SDW
phase
 is characterized by spin ordering with momentum
 $<S^z_Q> =1/2( <n_{\sigma\sigma}(Q)>-
<n_{\bar\sigma\bar\sigma}(Q)>)=\Delta_0$
and the CDW phase by the charge ordering. In the SCDW (OCDW)
the staggered magnetization (density) is equal to zero. Nevertheless,
the spin-current (charge-current) correlation functions
survive.

In our opinion for the case of high-$T_c$ cuprates it is the
interaction in the triplet channel which determines the behaviour
of the system and the nature of DW phase. From the theoretical
point of view it is this situation which corresponds to the
strong-coupling limit models : the Hubbard model and the $t-J$
model. For example for the latter with the $J$ term  written as
$H_J=\sum_{ij} J_{ij}\{a\vec{S}_i\vec{S}_j-(b/4)n_in_j\}$ one has
$V^{SDW}_{\bf q}=a J_{\bf q}$ while $V^{CDW}_{\bf q}=
-\frac{b}{4}J_{\bf q}$, i.e. the interaction in the triplet
channel is positive while in the singlet channel is negative. This
version is supported also by experiments in the high-$T_c$
cuprates : observed experimentally  (by INS, see for example
\cite{Rossat-Mignod} and NMR) strong magnetic response around
${\bf q}={\bf Q}$ is a phenomenological argument in a favor of a
strong momentum dependent interaction in a triplet channel, i.e.
of $V_{\bf q}=J_{\bf q}$ ($J>0$). However, we can not exclude an
importance of an interaction leading to the CDW (SCDW) order.


Another  point concerning the interaction
is its strength.
Depending on the ratio $|V|/W$ (where $W$ is
an energy bandwidth), maximal $T_{DW}^{max}$ can
be high or low. Respectively,  the
DW phase can lean out of SC state or can be
hidden under it. [In the presence of the
 interaction in the triplet channel,
$J_{\bf q}$, both  SDW and
SC instabilities occur around QCP1 under the same condition :
$J>0$, for the SC instability see \cite{On1}].
 It is temptating to identify the properties
obtained for the DW state with the properties observed
experimentally in the underdoped cuprates above $T_{sc}(\delta)$
and below $T^*(\delta)$. Indeed they have a striking resemblance,
as one can see when comparing
 Fig.\ref{f6} and Fig.\ref{f7}, Fig.\ref{f8}(a) and Fig.\ref{f8}(b) and  when comparing
 the
behaviour of the gap as a function of $Z$ (or doping,
$\delta_c-\delta$) with the experimental behaviour \cite{ARPES2}.
Our calculations (when considering both d-wave SC and DW
instabilities in the presence of interaction $J$ in the triplet
channel) show that the answer is quite subtle. When $t'/t=-0.2$
the ordered DW phase leans out of the SC phase for $t/J<1.90$, for
$t'/t=-0.3$ this happens when $t/J<1.55$. So far as realistic
value of $t/J$ for cuprates is estimated to be in the interval
$t/J = 1-3$, both variants (when the DW phase leans out  or is
hidden under the SC phase) are possible.

In this case two scenarios  can be discussed. First, the DW state
of s- or d- wave symmetry can coexist with d-wave SC state (this
situation is considered in \cite{Bouis}) resulting in appearance
of supplementary $\pi$-triplet state. Second, the DW state can be
suppressed under the SC state. Nevertheless, the strong
"fluctuation" memory of this ordering will affect on the behaviour
of disordered metallic state.

Even in the case if the long-range ordered DW phase is hidden
under SC phase it is this type of ordering which determines short
range correlations in the disordered metallic state
 above $T_c$ and
below $T^*(\delta)$. This point is discussed in
\cite{On2},\cite{PRL2}. The state below $T^*(\delta) \propto
\delta_c-\delta$ and above $T_c$ is quite exotic. It is almost
frozen in both temperature and doping. By this we mean that the
parameter $\kappa^2$ which determines a proximity to the ordered
DW phase does almost not change neither with $T$ nor with doping
remaining therefore quite low in a wide region in $T-\delta$ plane
below $T^*(\delta)$. Such a quasiordered state keeps a strong
memory about the ordered phase. Therefore,  electron properties in
this state should be close to those in the DW ordered state being
however characterized by strong damping. [By the way it is exactly
what is observed by ARPES. The experimental electron spectrum
 has a form shown in Fig.\ref{f7}. being however characterized
by a spectral function of a very damped form.
 Explicit consideration of the electron
spectrum in this state will be presented elsewhere].


Summarizing, we have studied the DW phase which is formed around
QCP1 (associated with ETT) and we have shown that this phase is
characterized by the following prominent features:

(i)  the specific  "flat" shape of the spectrum
in the vicinity of SP,

(ii) "disappearance" of the spectrum above some threshold value of
wavevector in the direction $(\pi,0)$ - $(\pi,\pi)$,

(iii) pseudogap in DOS with FL lying in it,

(iv) increasing of the
gap in the spectrum around SP and of the pseudogap in DOS
with decreasing doping for $\delta_c-\delta$

(v) angle dependence of the gap calculating from FL
which is of a d-wave type close to SP and flat close
to the direction $(1,1)$.

All these features have a striking similarity with the
experimental features revealed by ARPES in the
normal state of the underdoped hole-doped cuprates.


\section*{Appendix A}
The free energy density
in the approximation corresponding to considered in the paper is given by:
\begin{equation}
F=-T \frac{1}{N}\sum_{\bf k}\sum_{\alpha=1,2}
[\ln(2\cosh(\frac{\varepsilon_\alpha({\bf k},\Delta_{\bf k})}{2T})) +\frac{\Delta_{\bf k}^2}{4}]+
\mu n .
\label{mf}
\end{equation}
[Note that the equation (\ref{1.18})  corresponds to
$\partial F/\partial\Delta=0$.]
Therefore, the difference between free energies corresponding to
$\Delta=\Delta_1$ and $\Delta=\Delta_2$ is given by
\begin{equation}
F_{1}- F_{2} =\frac{\Delta_1^2-\Delta_2^2}{4V} -
T\frac{1}{N}\sum_{\bf k}
\sum_{\alpha=1,2}\ln\left(\frac{\displaystyle
\cosh(\frac{\varepsilon_{\alpha}({\bf k},\Delta_1)}{2T})}
{\cosh(\frac{\displaystyle\varepsilon_{\alpha}({\bf k},\Delta_2)}{2T})}
\right).
\label{mf12}
\end{equation}

One can check by numerical calculations that $F_1-F_2 <0$ for the
whole range of the coexistence of the two solutions.

Some analytical estimations can be also done for low $T$ based on the well-known expression \cite{AGD} for the difference
between thermodynamic potentials of the ordered and disordered states:
\begin{equation}
\delta\Omega=\Omega(\Delta_1) -\Omega(0) =\int_0^{\Delta_{1}}\frac{d(1/V)}{d\Delta}\Delta^2d\Delta.
\label{cor}
\end{equation}
 When substituting the expressions for $\Delta_1$ (\ref{DfZ}), (\ref{D0})
one gets
\begin{equation}
\delta F/t=\delta\Omega/t \sim -\frac{\sqrt{1-4|t'/t|^2}}{|t'/t|}
\frac{\Delta_1^3}{t^2\Delta(Z=0)} \sim -\frac{(Z_{cr}^{(1)})^3}{t^2\Delta(Z=0)}.
\label{o1}
\end{equation}
One can see that this correction is negative. Therefore, the solution $\Delta=\Delta_1$ is favorable with respect to the solution $\Delta=0$ for
any $Z_{cr}^{(1)}<Z<Z_{cr}^{(2)}$.

\bigskip
\noindent
{\it $^*$ Present address: Instit\"ut f\"ur Theoretische Physik, Universit\"at 
W\"urzburg,
D-97074 W\"urzburg, Germany}

\nopagebreak
\begin{onecolumn}

\begin{figure}
\vspace*{-0 mm}
\begin{center}
\epsfig{%
file=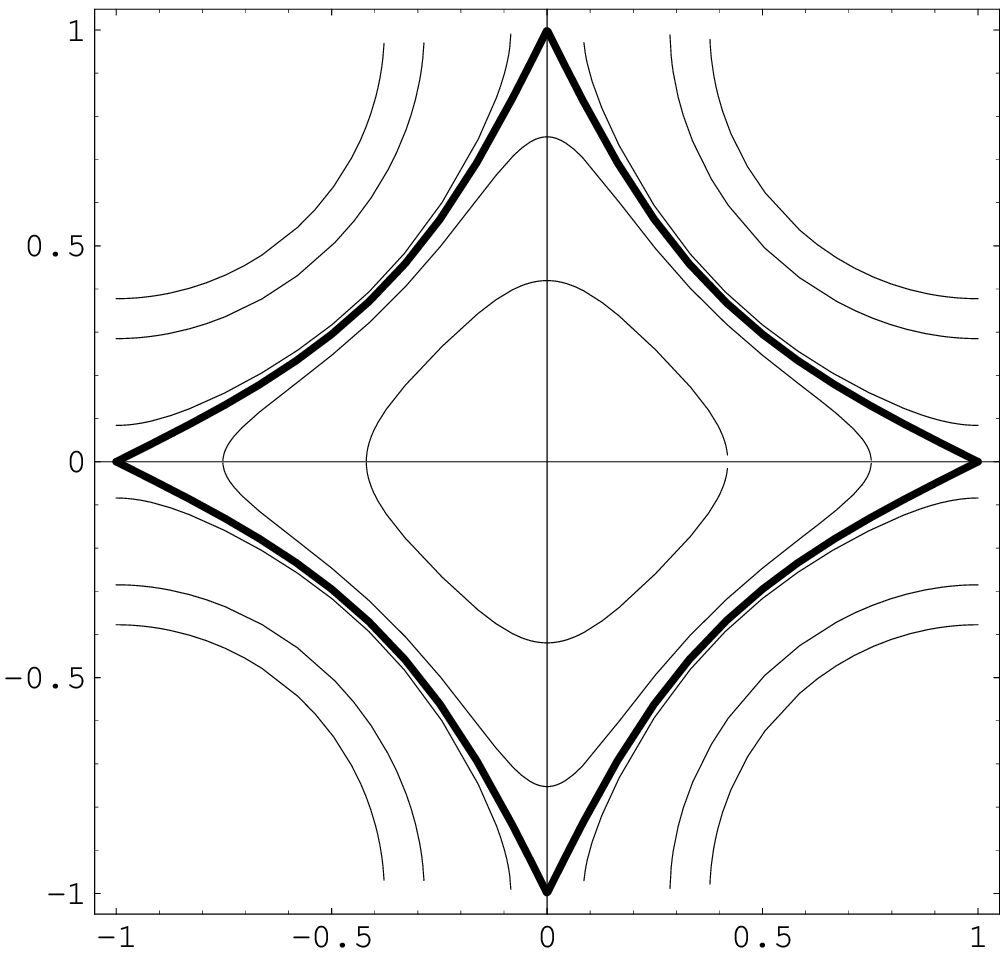,%
figure=fig1.eps,%
height=6cm,%
width=6cm,%
angle=0,%
}
\end{center}
\vspace*{-0 mm} \caption{ Fermi surface of the electron system
with the dispersion law (\ref{1}) for different Z and $t'/t=-0.3$.
The thick line corresponds to $Z=0$.} \label{f1}
\end{figure}

\begin{figure}
\vspace*{-3 cm}
\begin{center}
\epsfig{%
file=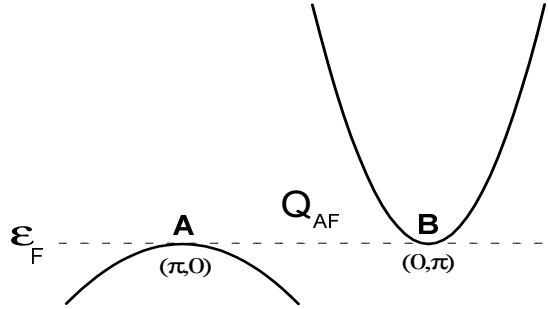,%
figure=fig2.eps,%
height=13cm,%
width=10cm,%
angle=0,%
}
\end{center}
\vspace*{-2.5 cm} \caption{ Schematic presentation of the electron
spectrum  in a vicinity of two SP's for $Z=0$} \label{f2}
\end{figure}

\begin{figure}
\begin{center}
\epsfig{%
file=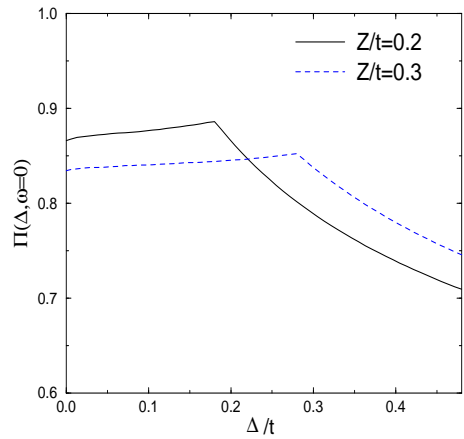,%
figure=fig3.eps,%
height=6cm,%
width=6cm,%
angle=-0,%
}
\end{center}
\caption{Calculated "polarization operator"
 $\Pi({\bf Q},Z,\Delta(Q))$
as a function of $\Delta_0$ for fixed $Z$ and $T=0$.} \label{f3}
\end{figure}

\begin{figure}
\begin{center}
\epsfig{%
file=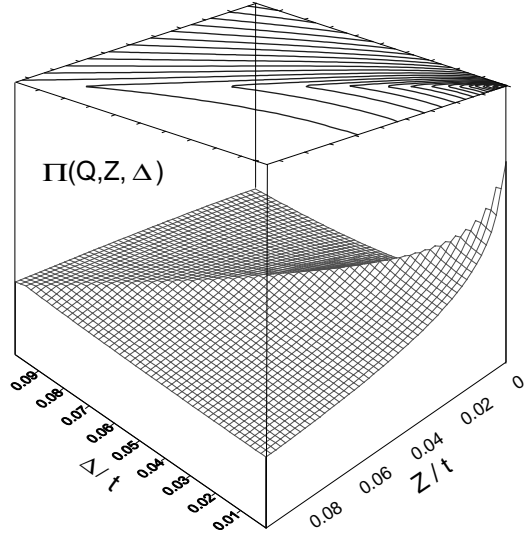,%
figure=fig4.ps,%
height=10cm,%
width=7cm,%
angle=0,%
}
\\
\vspace*{-1 cm}
\end{center}
\caption{
 $\Pi({\bf Q},Z,\Delta_0)$
as a function of $\Delta_0$ and $Z$ at $T=0$  and lines
$\Pi^{zz}({\bf Q})=const$ in $Z - \Delta_0$ coordinates.}\label{f4}
\end{figure}
\begin{figure}
\begin{center}
\epsfig{%
file=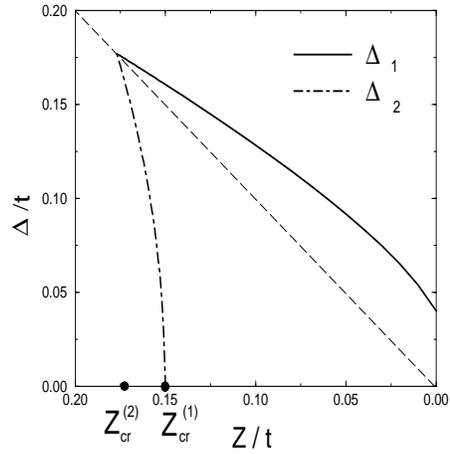,%
figure=fig5.eps,%
height=6cm,%
width=6cm,%
angle=0,%
}
\\
\mbox{}\\
\end{center} \caption{Gap $\Delta$ obtained by solving
equations (\ref{Pi}), (\ref{GE})
 as function
of Z (t/V=1.8, $t'/t=-0.3$). The solid line corresponds to
$\Delta_1(Z)$, the dot-dashed line to $\Delta_2(Z)$, the dashed
line to $\Delta_c(Z)$.} \label{f5a}
\end{figure}

\begin{figure}
\begin{center}
\epsfig{%
file=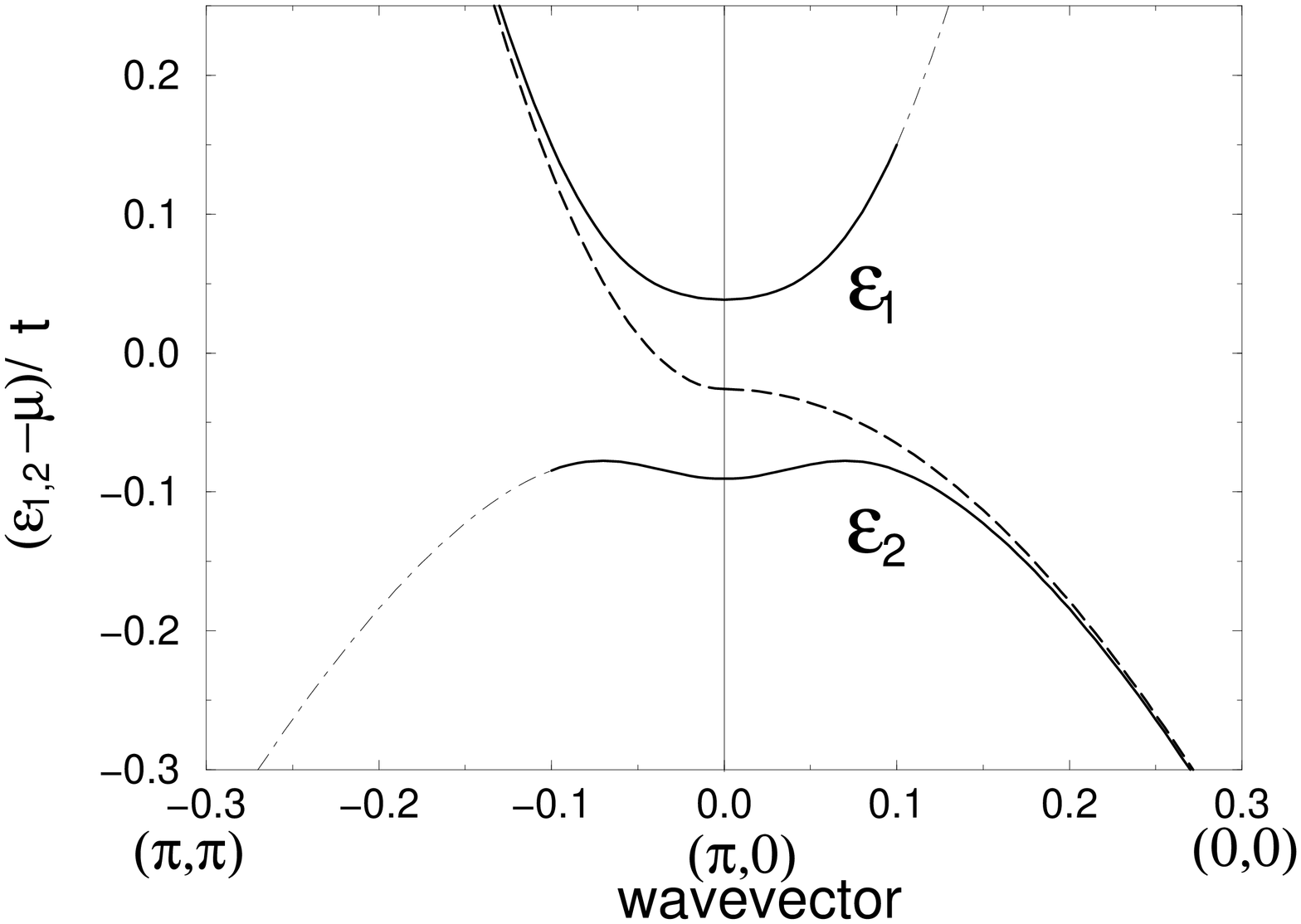,%
figure=fig6a.eps,%
height=8cm,%
width=5cm,%
angle=-90,%
}
\epsfig{%
file=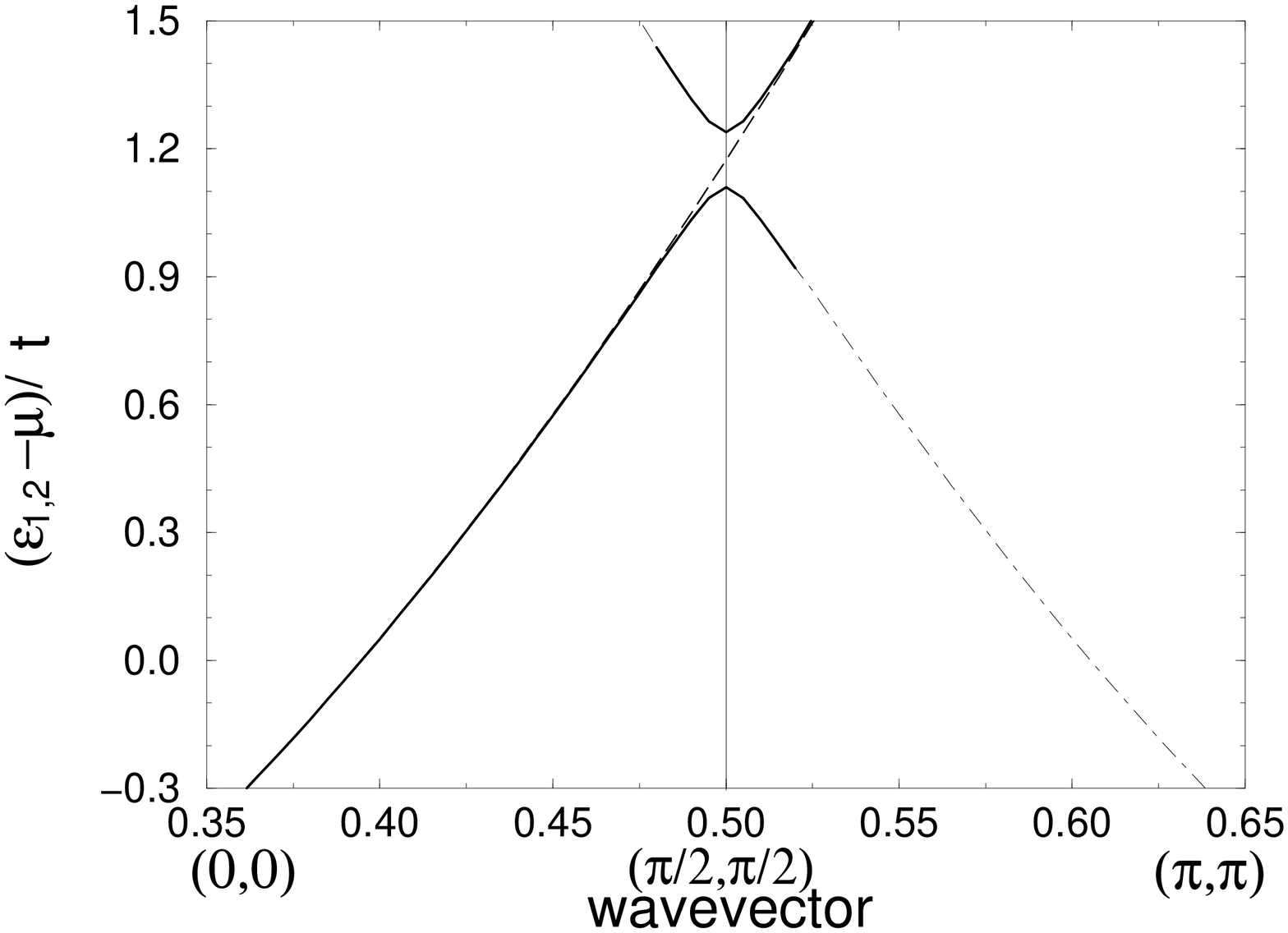,%
figure=fig6b.eps,%
height=8cm,%
width=5cm,%
angle=-90,%
}
\\
\vspace*{10mm} \hspace*{1cm}a)\hspace*{6cm}b)\\
\end{center}
\caption{One particle spectra along $(\pi,\pi) - (\pi,q_y/\pi)$
and $(\pi-q_x/\pi,0) - (0,0)$  symmetry lines (a) and in  $(0,0)$
-$(\pi,\pi)$ direction (b), $t'/t=-0.3, t/V=1.8, Z/t=0.03$. Long
dashed line is the bare spectrum, dot-dashed line corresponds to
the spectrum when the residue of the Green function (\ref{1.20})
less than 0.1.} \label{f6}
\end{figure}

\begin{figure}
\begin{center}
\epsfig{%
file=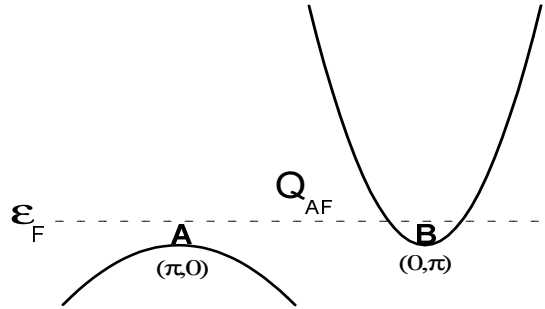,%
figure=fig7.eps,%
height=13cm,%
width=10cm,%
angle=0,%
}
\end{center}
\vspace*{-2.5cm} \caption{Schematic representation of the bare
spectrum in the vicinity of the two saddle points for $Z\ne 0$.}
\label{f5}
\end{figure}
\begin{figure}
\begin{center}
\epsfig{%
file=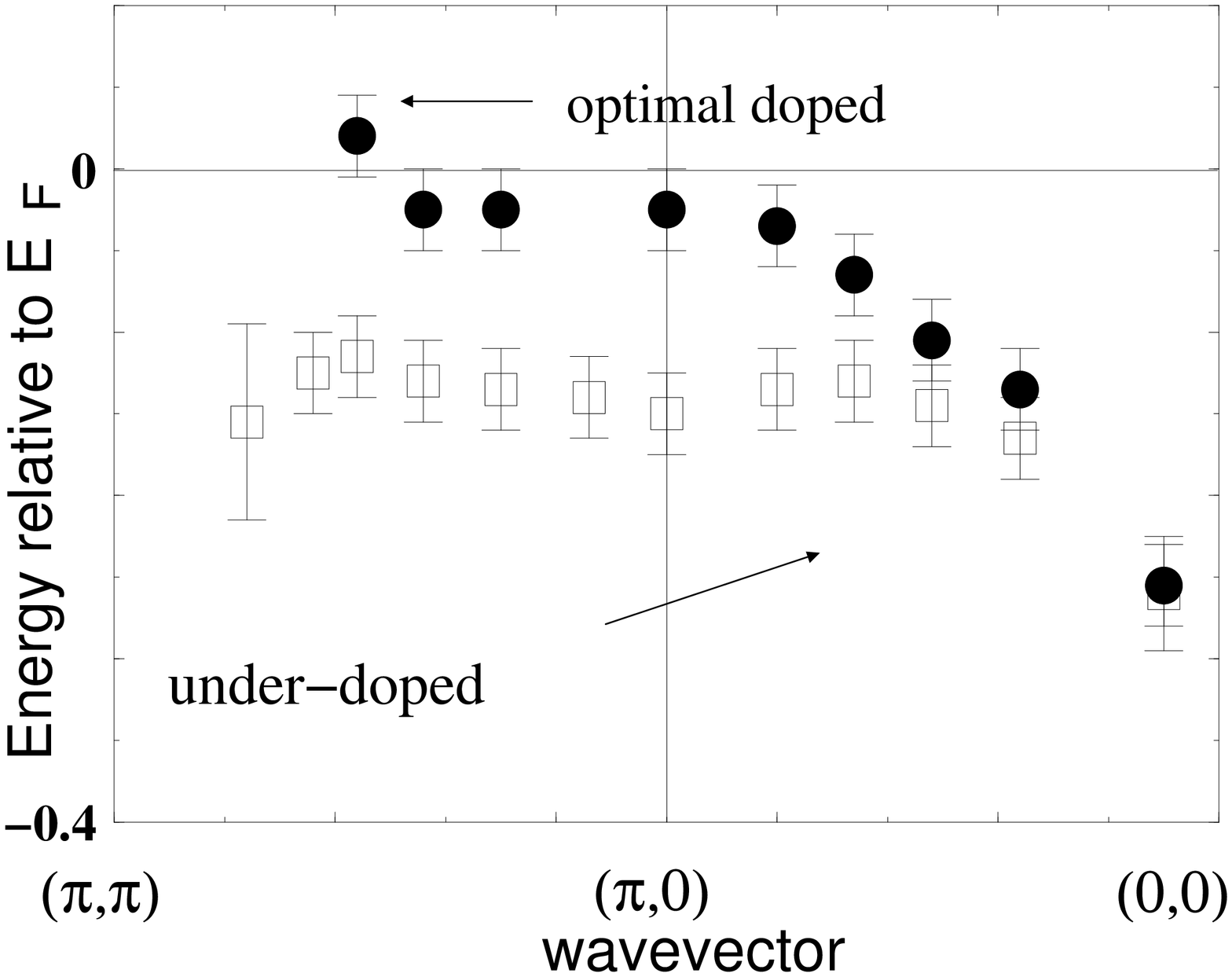,%
figure=fig8a.eps,%
height=8cm,%
width=5cm,%
angle=-90,%
}
\epsfig{%
file=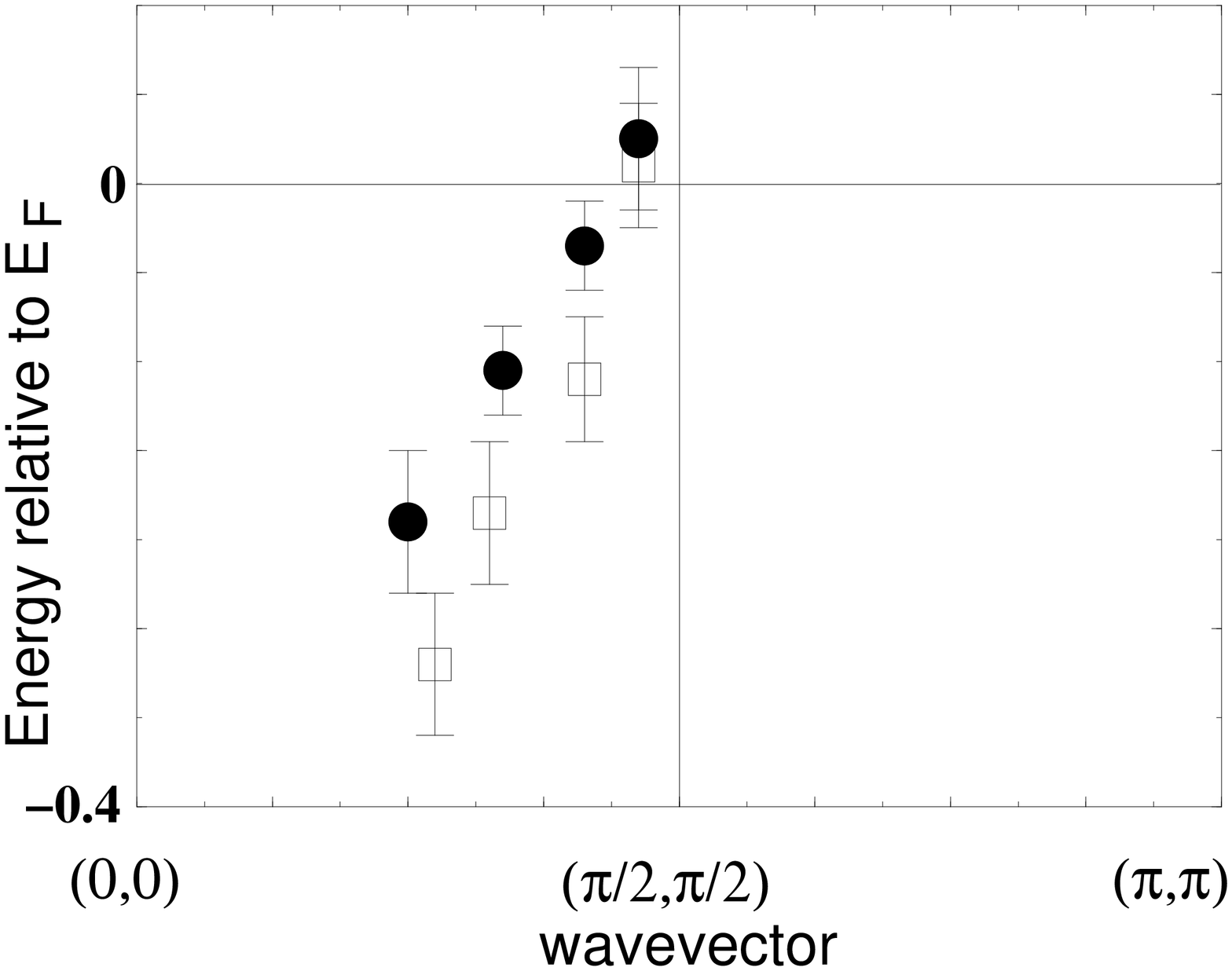,%
figure=fig8b.eps,%
height=8cm,%
width=5cm,%
angle=-90,%
}
\\
\vspace*{10mm} \hspace*{1cm}a)\hspace*{6cm}b)\\
\end{center}
\caption{Experimental one particle spectra along
$(\pi,\pi)-(\pi,0)-(0,0)$ symmetry lines (a) and in  $(0,0)$
-$(\pi,\pi)$ direction (b) measured in the overdoped regime of
BSCO. The data are taken from [10].} \label{f7}
\end{figure}

\begin{figure}
\begin{center}
\epsfig{%
file=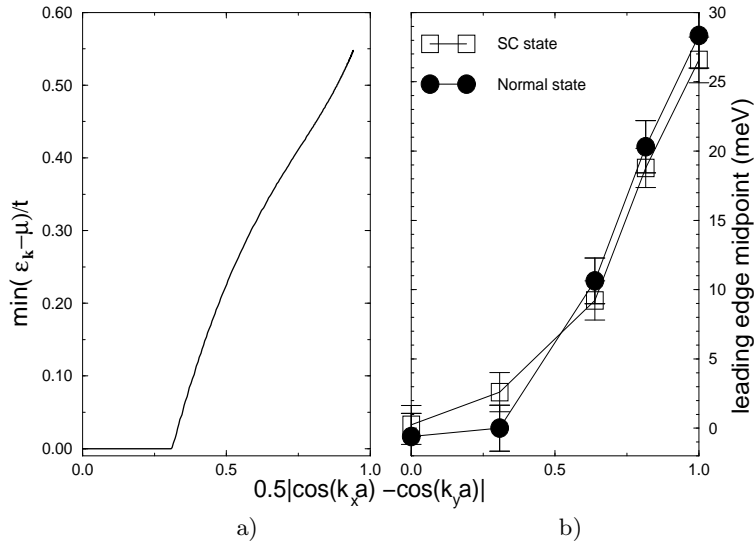,%
figure=fig9ab.eps,%
height=10cm,%
width=7cm,%
angle=-90,%
}
\\
\vspace*{-0mm} \hspace*{1cm}a)\hspace*{4cm}b)\\
\end{center}
\caption{Theoretical angle dependence of the SDW gap 
 calculated from FL in the underdoped
regime $Z>0$
 ($t'/t=-0.3$, $t/V=1.7$, $Z/t=0.3$) (a) and
the experimental leading edge midpoint measured by ARPES in the
underdoped BSCO [11] (b).} \label{f8}
\end{figure}

\begin{figure}
\begin{center}
\epsfig{%
file=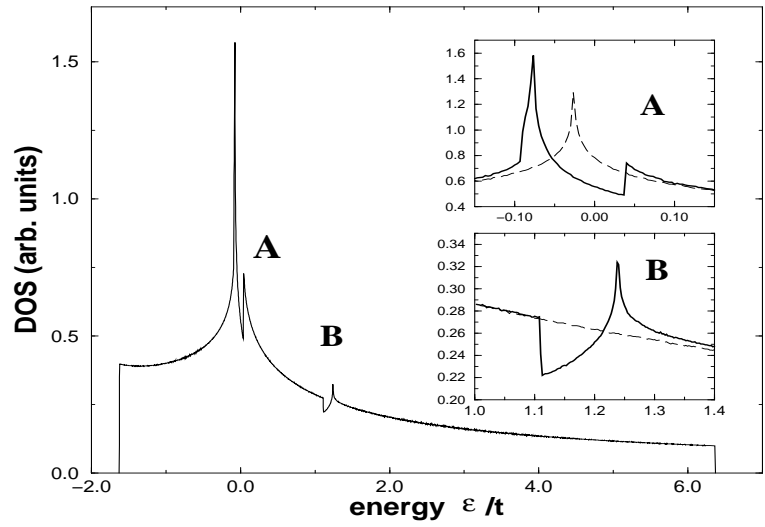,%
figure=fig10.eps,%
height=10cm,%
width=7cm,%
angle=-90,%
}
\end{center}
\vspace*{5mm} \caption{ Density of states in the ordered
"excitonic" phase calculated for $Z/t=0.03$ ($t/V=1.8,
t'/t=-0.3$).
 Dashed
line corresponds to the DOS in the initial metallic state.}
\label{f9}
\end{figure}

\begin{figure}
\begin{center}
\epsfig{%
file=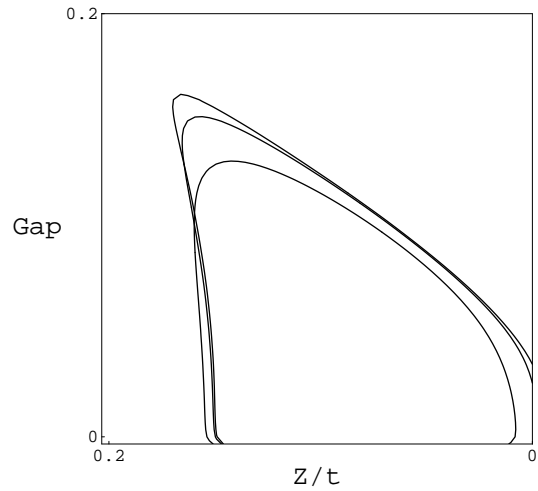,%
figure=fig11.eps,%
height=7cm,%
width=7cm,%
angle=-0,%
}
\end{center}
\caption{The DW gap in $t$ units as a function of $Z$ for
increasing temperature : $T/t=0.005, 0.1, 0.2$ ($t'/t=-0.3$,
$t/V=1.8$).} \label{f10}
\end{figure}
\begin{figure}
\begin{center}
\epsfig{%
file=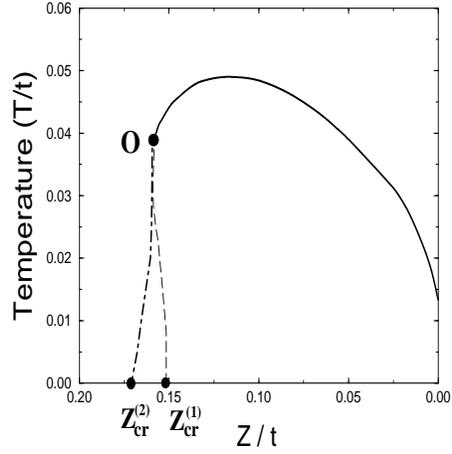,%
figure=fig12.eps,%
height=6cm,%
width=6cm,%
angle=0,%
}
\\
\vspace*{-0mm}
\end{center}
\caption{ Phase diagram around QCP1 in T-Z coordinates ($t/V=1.8$,
$t'/t=-0.3$). We show only the regime $Z>0$ corresponding to the
anomalous behavior. The solid line is a line of second-order phase
transition, the dot-dashed is a line of first-order phase
transition and the dashed line is a line of instability of the
disordered metal state (spinodal). The point $O$ is a tricritical
point.} \label{f11}
\end{figure}

\begin{figure}
\begin{center}
\epsfig{%
file=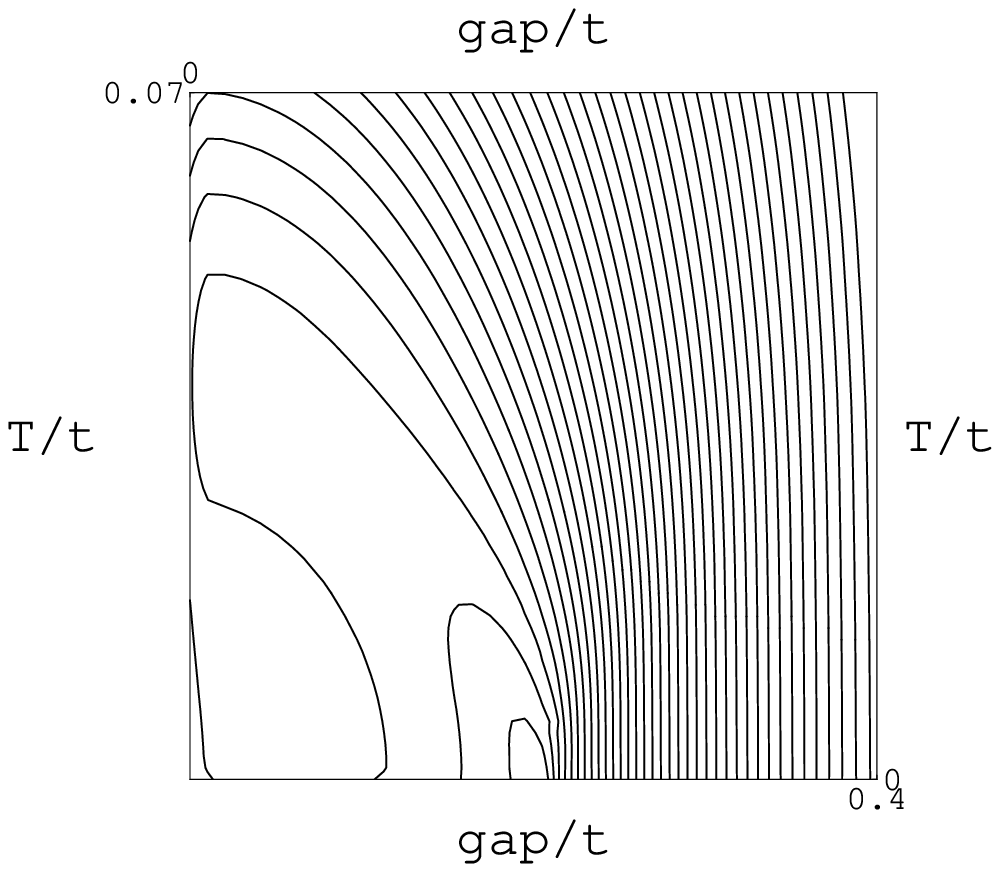,%
figure=fig13a.eps,%
height=7cm,%
width=7cm,%
angle=-0,%
}
\epsfig{%
file=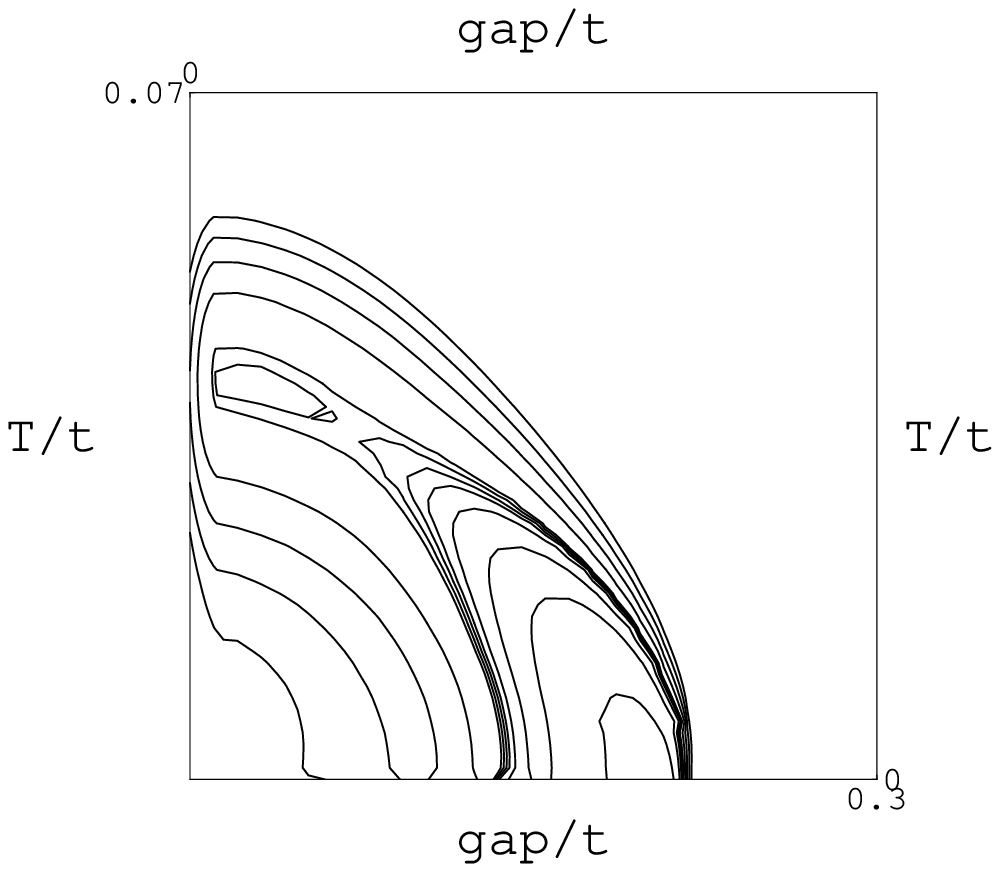,%
figure=fig13b.eps,%
height=7cm,%
width=7cm,%
angle=-0,%
}
\\
\hspace*{1cm}a)\hspace*{6cm}b)\\
\end{center}
\caption{Lines of $\Pi({\bf Q},Z,\Delta)=const$ for fixed
$t/V=1.8$ and different  $Z$. The plot (b) is a zoom of the plot
(a) corresponding to the coexistence of the two solutions for the
gap.} \label{f13}
\end{figure}

\begin{figure}
\begin{center}
\epsfig{%
file=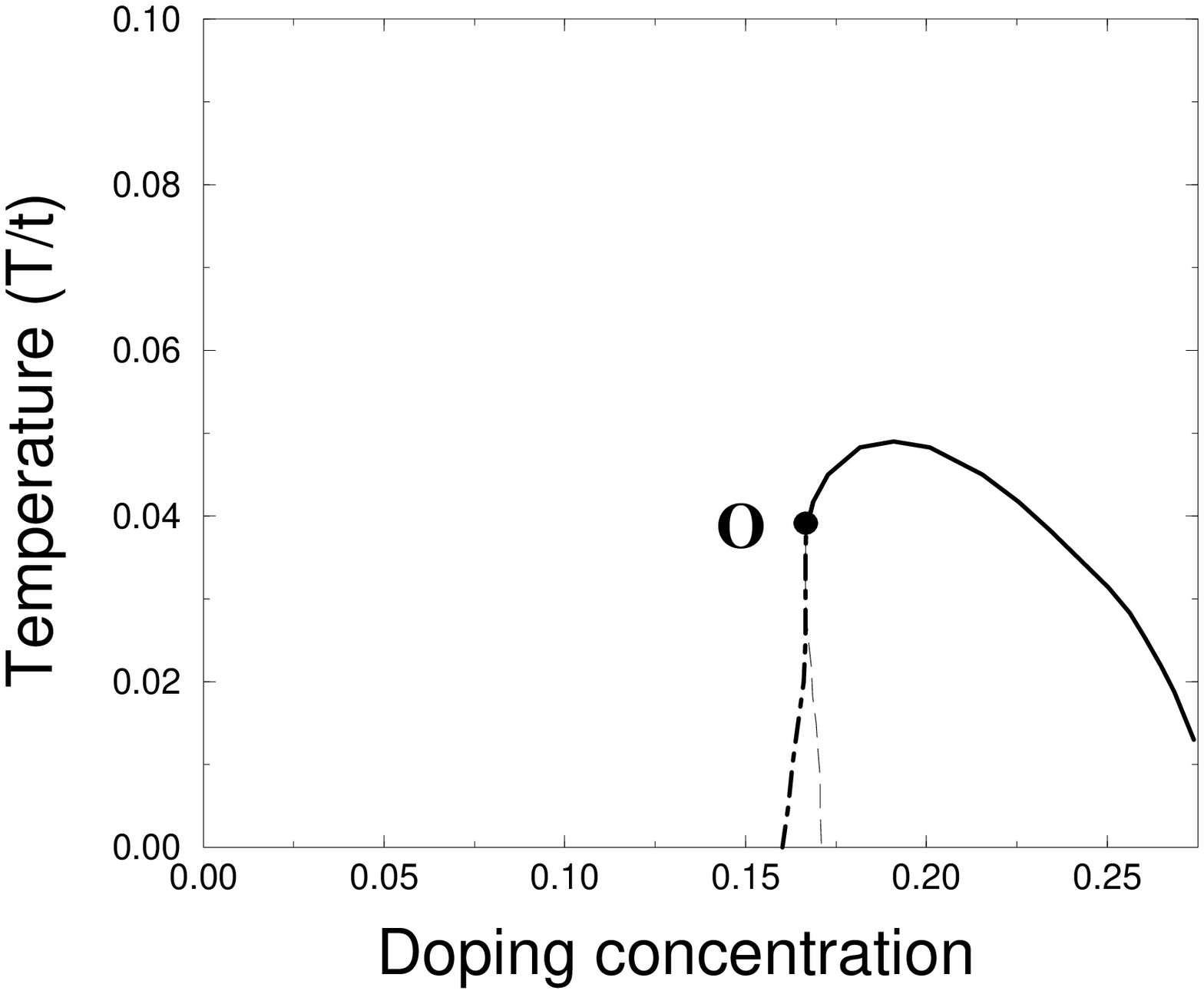,%
figure=fig14.eps,%
height=6cm,%
width=6cm,%
angle=0,%
}
\\
\vspace*{-0mm}
\end{center}
\caption{ Phase diagram around QCP1 in T-$\delta$ coordinates
($t/V=1.8$, $t'/t=-0.3$)} \label{f12}
\end{figure}
\end{onecolumn}

\begin{thebibliography}{99}
\bibitem{DW} We write down the expression for SDW and OCDW
polarization operator just as an example. The pair
of equations for CDW and SCDW is different only due to definition
of $\Pi^{DW}$.
\bibitem{const} This is true under the condition of $|t'/t|$ not too small.
\end{thebibliography}
\end{document}